\documentclass[12pt]{iopart}
\usepackage{graphicx}

\begin{document}

\title[N. D. Lai et al., Single nitrogen-vacancy color center in diamond nanocrystal]{Optical determination and magnetic manipulation of single nitrogen-vacancy color center in diamond nanocrystal}

\author{Ngoc Diep Lai$^{1}$, Dingwei Zheng$^{1, 2}$, Fran\c cois Treussart$^{1}$, \\ and Jean-Fran\c cois Roch$^{1}$}

\address{$^{1}$
Laboratoire de Photonique Quantique et Mol\'eculaire, UMR CNRS 8537, \\Ecole Normale Sup\'erieure de Cachan, 61 av du Pr\'esident Wilson, 94235 Cachan, France.}
\address{$^{2}$
State Key Laboratory of Precision Spectroscopy, East China Normal University, \\3663 Zhongshan Road North, Shanghai 200062, China.}
\ead{ndlai@lpqm.ens-cachan.fr}

\begin{abstract}
The controlled and coherent manipulation of individual quantum systems is a fundamental key for the development of quantum information processing. The nitrogen-vacancy (NV) color center in diamond is a promising system since its photoluminescence is perfectly stable at room temperature and its electron spin can be optically read-out at the individual level. We review here the experiments currently realized in our laboratory, concerning the use of single NV color center as single photon source and the coherent magnetic manipulation of the electron spin associated to a single NV color center. Furthermore, we demonstrate a nanoscopy experiment based on saturation absorption effect, which allows to optically pin-point single NV color center at a sub-$\lambda$ resolution. This opens a possibility to independently address two or multiple magnetically-coupled single NV color centers, which is a necessary step toward the realization of a diamond-based quantum computer.

\end{abstract}

\submitto{\ Journal of Advances in Natural Sciences}
\maketitle

\section{Introduction}
The nitrogen-vacancy (NV) color center in diamond has been identified as one system of choice for single photon generation [1] and individual electron spin manipulation [2] at room temperature. In particular, its electron spin has an exceptional long coherence time, which is advantageous over other system envisioned for quantum computation operating at room temperature [3] and also for measuring magnetic field with high sensitivity and nanoscale resolution [4]. To this aim, a diamond nanocrystal (nanodiamond) containing a single NV color center is a key element, because it can be manipulated in space with a high precision. On the other hand, the nanodiamond containing a NV color center can also be coupled to a photonic system such as photonic crystal [5] resulting in an enhancement of the single photon emission. However, for most of these applications, it is important to know the orientation of the optical dipoles  of the color center embedded in the nanodiamond  in order  to efficiently couple the NV photoluminescence to the photonic structure. Such determination is directly related to spin orientation measurement since it has been already shown that the electron spin is orthogonal to the two independant optical dipoles involved in the photoluminescence. Spin orientation knowledge is useful to build-up a vectorial ultra-sensitivity NV single-spin based magnetometer. Besides, it is also useful to be able to address optically and individually a single NV color center separated spatially from others at nanometer scale. The study of two or multiple magnetically coupled NV electron spins will be the first step toward the realization of diamond-based quantum computer operating at room temperature.

In this article, we first present the use of a single NV color center in nanodiamonds as an efficient single photon source at room temperature. Secondly, we demonstrate a simple technique based on the optically detected magnetic resonance to determine precisely the orientation of a single electron spin associated to a single NV color center embedded in an arbitrary nanodiamond. We finally demonstrate a microscopy experiment based on saturation absorption effect, which allows to optically pin-point single NV color centers at a sub-$\lambda$ resolution. 

\section{Experimental Setup}

The nanodiamonds sample is prepared following a procedure described in Ref. 1, starting from type Ib synthetic diamond powder (de Beers, Netherland). The nanocrystals (about $90$~nm mean diameter) are spin-coated on a dielectric mirror or on a glass substrate. The orientation of a nanodiamond crystallographic axis relative to the substrate surface is randomly distributed. Due to the $C_{\rm 3v}$ symmetry of the NV-centers in the diamond crystal, a single NV-center is aligned along one of four possible orientations [6,7] associated to the [111] axis of the crystal, which is un-determined due to the arbitrary orientation of the nanodiamond.

Optical excitation and detection of the NV color center photoluminescence is realized with a home-built setup, illustrated in Fig. 1(a). The NV color center is excited with a cw laser (wavelength = 532 nm) and emits a broadband photoluminescence (about 100 nm - FWHM spectral width) centered at 670 nm. The excitation beam is tightly focused on the sample, using an oil immersion high numerical aperture microscope objective (Obj). The focus point is raster scanned relative to the sample with nanometer resolution, using a 3D-piezoelectric translation system. The photoluminescence of the excited NV-center is collected by the same objective and spectrally filtered from the remaining pump light by a long-pass filter with wavelength cut-off at 580 nm. A standard confocal detection system (not shown) is also used to select the luminescence coming from a sample volume of about 1~$\mu$m$^3$, ensuring the collection of light emitted by only one nanocrystal. The photoluminescence signal is finally detected by a silicon avalanche photodiode (APD) operating in the single-photon counting regime. 

To apply the microwave field for the electron spin resonance (ESR), the sample is placed on a circuit board with a strip line for microwave field input-output connection. A 20 $\mu$m-diameter copper wire placed over the sample is soldered to the strip lines. The wire is positioned within 20~$\mu$m from the optically addressed NV color center. The microwave power injected into the strip lines is fixed at 25 dBm for all measurements. The ESR measurement is then realized by sweeping the microwave frequency with a step of 0.3~MHz/channel, while detecting corresponding luminescence photons. The external static magnetic field (\textbf{B}) applied to the NV center can be varied in magnitude and in orientation by moving and rotating a permanent magnet, which is mounted on three-axis translation and rotation stages, with respect to the sample.

In order to image the NV color centers with sub-$\lambda$ resolution, the green excitation beam is passed through a phase mask (vortex phase plate), which allows obtaining a doughnut spot at the focusing region of the microscope objective. The phase mask position is controlled precisely by using a two-axis translation stage. The use of this phase mask necessites a high excitation power and it is used only for sub-$\lambda$ imaging experiment.

\begin{figure}
\centering
\includegraphics[width=11 cm]{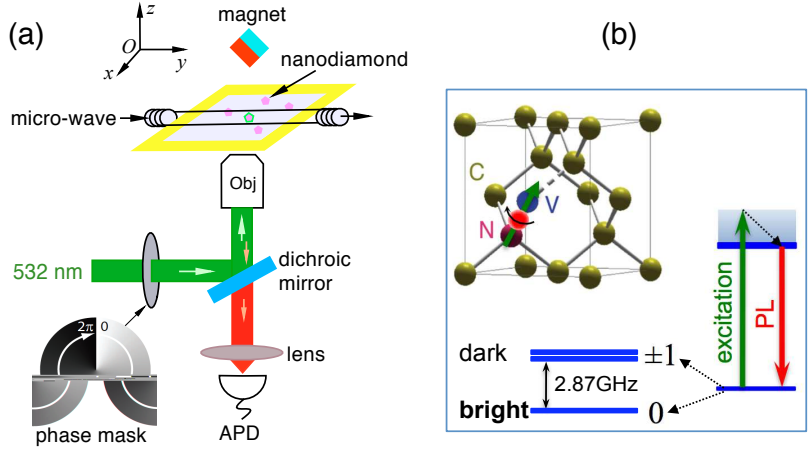}
\caption{(a) Experimental setup used to i) address individual color center, ii) manipulate single electron spin associated to single NV color center, and iii) image NV color centers in diamond nanocrystals with sub-$\lambda$ resolution. (b) Representative structure of NV color center in diamond matrix and schematic diagram of the NV ground-state electron spin energy levels.}
\end{figure}

\section{Generation of single photon from single NV color center in a diamond nanocrystal} 
Single NV color center in diamond stands as one of the promising candidates among different kinds of solid-state single-photon sources due to their high efficiency and perfect photostability. Consisting of a substitutional nitrogen atom and a vacancy in an adjacent lattice site, NV color center absorbs well an excitation light at 532 nm-wavelength and emits a broadband photoluminescence with a maximum at around 670 nm and with a zero-phonon line at 637 nm. This color center plays a role as an atom or a molecule, which absorbs a single photon and emits only one photon at a time. 

In order to realize such a single photon source, the sample is first scanned over a large area. A photoluminescence image is obtained as shown in Fig. 2(a), in which each bright spot represents the photoluminescence of NV color centers embedded in one nanodiamond. We then reposition the excitation beam and excite only one nanodiamond. To know if the nanodiamond contains only a single NV color center, the emitted photoluminesence is sent to a so-called Hanbury Brown and Twiss system, which separates a photoluminescence beam into two beams and detected by two APDs (Fig. 2(b)). If the emission is single photon, no coincidence detection can be obtained because one photon cannot be splitted into two. Some of the nanodiamonds, which are numbered C3, C4 and C5 in Fig. 2(a), therefore are demonstrated to possess only a single NV color center, as indicated by the zero-coincidence detection by the two APDs shown in Fig. 2(c), at zero delay.  A single photon source at room temperature is then obtained by exciting a well determined nanodiamond. A triggered single-photon source was also obtained [1] by using a pulsed excitation of a single NV color center in a nanodiamond.

Note that for the ESR experiment, a single photon emission is always verified before all ESR measurements in order to magnetically manipulate a single electron spin associated to a single NV center.
 
\begin{figure}
\centering
\includegraphics[width=14 cm]{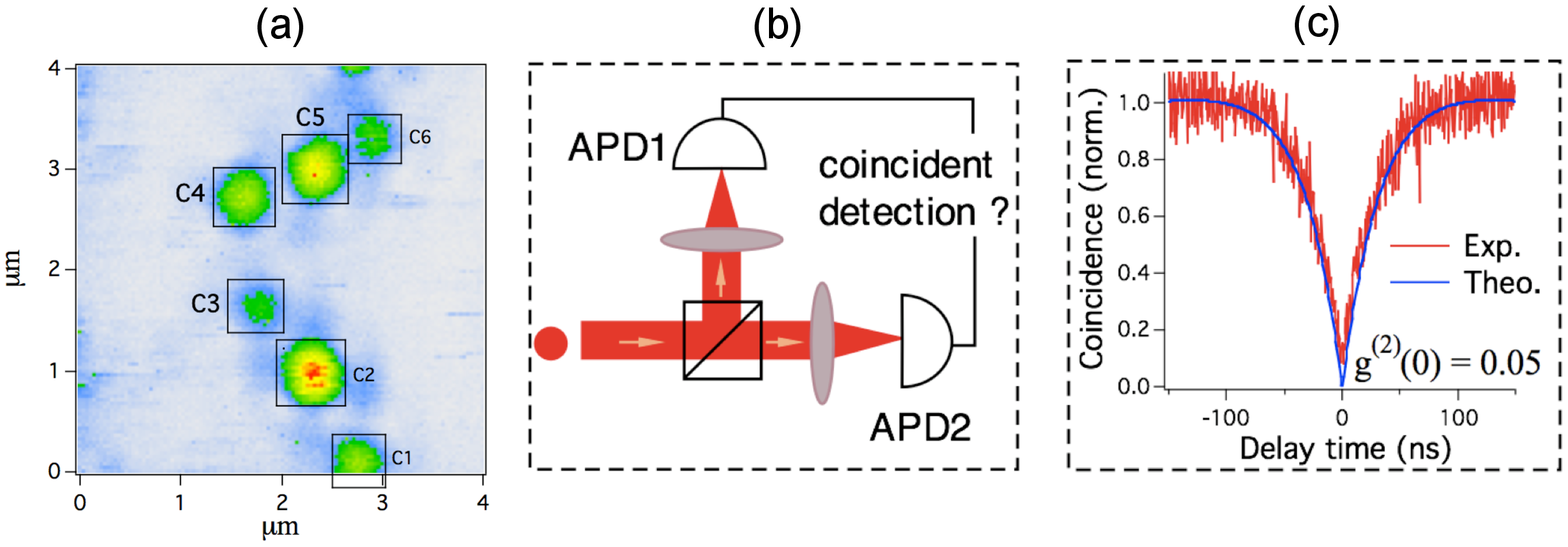}
\caption{(a) Photoluminescence image of NV color centers in diamond nanocrystals. (b) The Hanbury Brown and Twiss (HBT) setup used to verify single photon emission. (c) Time intensity correlation histogram of NV-center photoluminescence shows photon antibunching at zero delay, which is  the signature of single-photon emission from an individual NV color center.}
\end{figure}

\section{Optically detected magnetic resonance of a single NV color center electron spin}

In our sample, diamond nanocrystals mostly contain negatively charged NV$^-$ color centers. The ground state ($^3A_2$) of the NV$^-$-center is known to have an electron spin triplet structure with a zero-field splitting of 2.87 GHz between the $m_S = 0$ and the degenerate $m_S = \pm1$ states [6,7]. The continuous laser excitation at 532 nm-wavelength optically prepares the electron spin into the brighter $m_S = 0$ state (see Fig. 1(b)). When the microwave is applied to the NV$^-$-center and its frequency is resonant with one of the spin transitions $m_S = 0 \leftrightarrow m_S = \pm1$, the population is redistributed between the two levels, and the photoluminescence intensity decreases, as shown in Fig 3(a). The determination of the spin resonance via the modulation of the photoluminescence intensity is therefore called optically detected magnetic resonance (ODMR).

We now apply this ODMR technique to determine precisely the orientation of a single NV color center embedded in an arbitrarily oriented nanodiamond, relative to the laboratory reference frame. Firstly, we apply the \textbf{B}-field along the $Oz$-axis (vertical), and we change \textbf{B}-magnitude by translating the permanent magnet along this $Oz$-axis (see Fig. 1(a) and Fig. 3(c)). By scanning the microwave frequencies, we obtained two resonance peaks for each magnet position (Fig. 3(b)), due to the Zeeman effect, with an increasing splitting upon the increase of the \textbf{B}-field magnitude (Fig. 3(c)). By comparing with the numerical simulation using equation governing the spin Hamiltonian of the system, we infer the $\theta$-angle ($57^{\circ}$) between the \textbf{B}-field and the NV spin orientation. With only $\theta$-angle, the NV spin can be in any direction belonging to a cone with $Oz$ as its symmetrical axis.

We then keep the distance between the magnet and the NV-center constant in order to keep constant the \textbf{B}-magnitude applied to the NV spin. We rotate the magnet around the  $Oz$-axis, by an angle $\phi$, with the condition that the magnet North-South axis is always pointing toward the NV-center (Fig. 3(d)). This results in a change of the angle between \textbf{B}-field and spin axis. According to Zeeman effect, the resonance frequencies, corresponding to $m_S = 0 \leftrightarrow m_S = -1$ and $m_S = 0 \leftrightarrow m_S = +1$ transitions, vary as a function of $\phi$. When the angle between the \textbf{B}-field and the spin axis angle is minimal (corresponding to the $\phi$ = $105^{\circ}$, azimuthal angle), the splitting between the two resonance peaks is maximal. 

By realizing these two measurements, we determined accurately in the laboratory frame the orientation of a single electron spin associated to a single NV color center in an arbitrary diamond nanocrystal [7].

\begin{figure}
\centering
\includegraphics[width=14 cm]{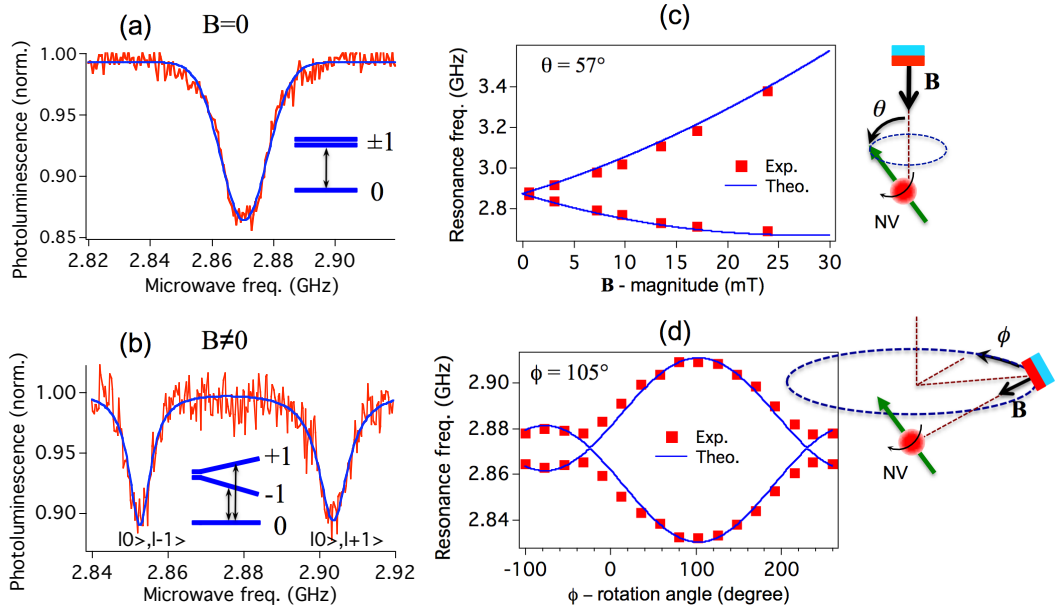}
\caption{Coherent manipulation of single electron spin associated to a single NV color center in a diamond nanocrystal. Electron spin resonance spectrum obtained optically with B = 0 mT (a) and with B $\neq$ 0 mT (b). (c), (d) Determination of the orientation of a single NV spin by the ODMR method: (c) First measurement: $\theta$-angle is obtained by changing the magnetic field magnitude, which is controlled by translating vertically the magnet along the $Oz$-axis; (d) Second measurement: $\phi$-angle is obtained by changing the magnetic field orientation. The magnet is rotated horizontally around the $Oz$-axis.}
\end{figure}

\section{Sub-diffraction-limited optical imaging of diamond nanocrystals}  

For NV color centers in diamond, the photoluminescence emission is saturated at rather low excitation power (about 1 mW, corresponding to a saturation intensity of about 105~W/cm$^2$), \textit{i.e.}, the emission rate does not increase for the excitation power above 1~mW. By using a doughnut-shaped excitation beam, the saturation is only reached in the doughnut region while the doughnut center remains dark,  \textit{i.e.}, there is no excitation and no emission at the doughnut center. The size of the dark spot can become nanometric if the excitation power is very high compared to the excitation saturation level. By scanning the diamond nanocrystals sample using the doughnut beam, we could therefore determine precisely the position of each NV color center, which is represented in photoluminescence image by a dark spot.

Figure 4 shows the photoluminescence images of NV color centers, obtained with a gaussian excitation beam (standard confocal microscope) and with a doughnut excitation beam (STED-like microscope [8]). It is difficult to distinguish the position of NV color centers using a commonly used confocal technique. In contrast, when using the doughnut-shaped excitation beam, each NV color center is represented by a dark spot and distinguished from the others. Moreover, with an excitation power of 13 mW, the dark spot becomes effectively nanometric in size, and two NV color centers separated by only about 250 nm are clearly identified (Fig. 4(f)). This imaging technique is effectively very useful to determine the position of NV color centers, with a precision well below the diffraction limit of regular confocal microscopy.

In order to achieve an even higher resolution, it is necessary to improve the quality of the doughnut beam, \textit{i.e.}, optimize the phase mask for the excitation wavelength. Besides, to address optically a single color center without perturbing the neighbours and to obtain directly the photoluminescence from nanospot, the realization of a so-called positive STED will be neccessary, by considering for example two alternative excitations, one has a gaussian form and another has doughnut form [8]. The success of this imaging technique will open the way to single spin optical addressing in the realization of diamond-based quantum computer functioning at room temperature.

\begin{figure}
\centering
\includegraphics[width=14 cm]{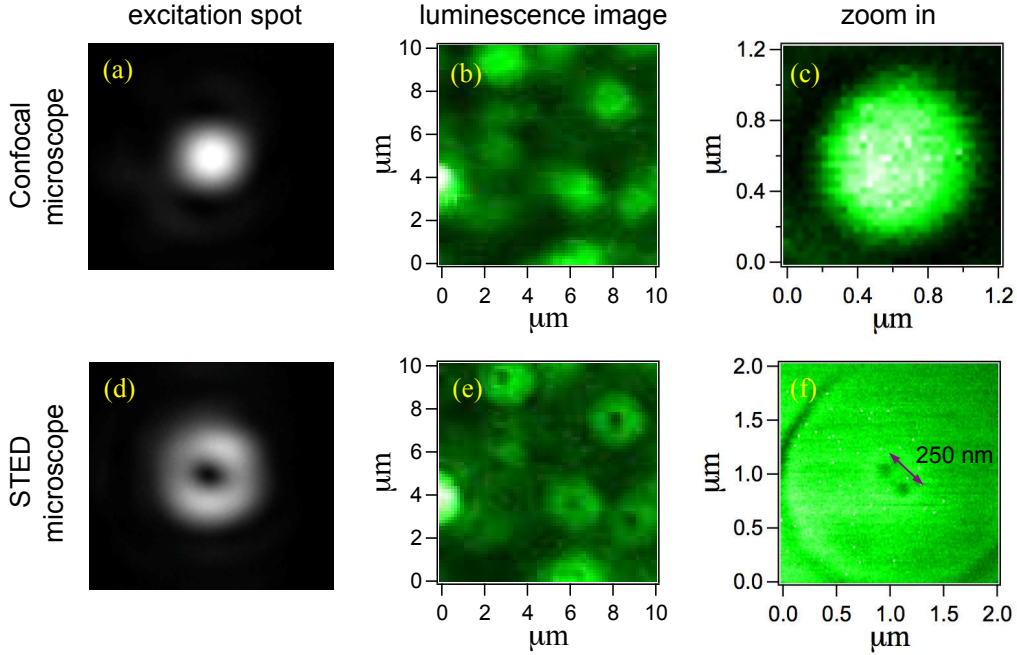}
\caption{Comparison of luminescence images obtained by a commonly used confocal microscope and a negative Stimulated Emission Depletion (STED) microscope. (a), (d) Focusing spots of excitation beam. (b), (e) Large-area photoluminescence images. (c), (f) Zoom in photoluminescence image, obtained at 13 mW excitation power. The FWHM of the dark spot is about 40 nm, which is much smaller than the diffraction limit. Two NV color centers, separated by 250 nm are therefore clearly identified by STED technique while they are un-separated with a confocal microscope.}
\end{figure}

\section{Conclusion}  

Our work focused on the study of fundamental physical properties of NV color centers in diamond nanocrystals and applications based on its optical and magnetic properties. The NV color center is demonstrated to be an excellent single photon source, reliable and stable at room temperature, which can be used for many applications, such as quantum experiment or quantum key distribution. The single electron spin associated to a single NV color center is coherently and optically manipulated and it becomes now an excellent candidate for quantum information processing, in particular for building up a new ultrasensitive magnetometer. A super-resolution imaging technique is also demonstrated, based on the use of a doughnut-shaped excitation beam at high excitation power. This allows to perform optical addressing of individual spins in a future experiment of coupled multiple electron spins, which pave the way to a room diamond-based quantum computer. 

\section*{Acknowledgments}
This work is supported by the European Commission through EQUIND (FP6 project number IST-034368) and NEDQIT (ERANET Nano-Sci) projects, by the National Research Agency through PROSPIQ (ANR-06-NANO-041-01) project, and by the RTRA ``Triangle de la Physique'' (B-DIAMANT project).


\end{document}